\documentclass[3p]{elsarticle}

\usepackage{a4wide}
\usepackage{float}
\usepackage{amsmath}
\usepackage{amssymb}
\usepackage{graphicx}
\usepackage[latin1]{inputenc} 
\usepackage{rotating}
\usepackage{setspace}
\usepackage{subfigure}
\usepackage{sidecap}
\usepackage{upgreek}
\usepackage{microtype}
\usepackage{lineno}
\usepackage{url}
\linenumbers

\begin{document}
\begin{frontmatter}


\title{Study of the accumulation layer and charge losses at the Si-SiO$_2$ interface in p$^+$n-silicon strip sensors.}


\renewcommand{\thefootnote}{\fnsymbol{footnote}}

 \author[]{Thomas Poehlsen$^{a,}$\corref{cor1} }
 \author[]{Julian Becker$^b$}
 \author[]{Eckhart Fretwurst$^a$}
 \author[]{Robert Klanner$^a$}
 \author[]{Jörn~Schwandt$^a$}
 \author[]{Jiaguo~Zhang$^a$}
 \cortext[cor1]{Corresponding author; Email: thomas.poehlsen@desy.de.}

 \address{$^a$Institute for Experimental Physics, University of Hamburg, Hamburg, Germany}
 \address{$^b$DESY, Hamburg, Germany}

 \begin{abstract}

 Using the multi-channel Transient Current Technique the currents induced by electron-hole pairs, produced by a focussed sub-nanosecond laser of 660~nm wavelength close to the Si-SiO$_2$ interface of $p^+n$ silicon strip sensors have been measured, and the charge-collection efficiency determined.
  The laser has been operated in burst mode, with bursts typically spaced by 1~ms, each consisting of 30~pulses separated by 50~ns.
 In a previous paper it has been reported that, depending on X-ray-radiation damage, biasing history and humidity, situations without charge losses, with hole losses, and with electron losses have been observed.
  In this paper we show for sensors before and after irradiation by X-rays to 1~MGy~(SiO$_2$), how the charge losses change with the number of electron-hole pairs generated by each laser pulse, and the time interval between the laser pulses.
 This allows us to estimate how many additional charges in the accumulation layers at the Si-SiO$_2$ interface have to be trapped to significantly change the local electric field, as well as the time it takes that the accumulation layer and the electric field return to the steady-state situation.
  In addition, results are presented on the change of the pulse shape caused by the plasma effect for high charge densities deposited close to the Si-SiO$_2$ interface.

\end{abstract}





 \begin{keyword}
 Silicon strip sensors \sep X-ray-radiation damage  \sep charge losses \sep Si-SiO$_2$ interface \sep accumulation layer \sep plasma effect  \sep XFEL


 \end{keyword}

 \end{frontmatter}



 \section{Introduction}
  \label{chapter:introduction}
 The high instantaneous intensity and the 4.5 MHz repetition rate of the European X-Ray Free-Electron Laser (XFEL) \cite{XFEL, XFEL2, Tschentscher:2011} pose new challenges for imaging X-ray detectors~\cite{Graafsma:2009, Klanner:2011}.
 The specific requirements for the detectors include a dynamic range of 0, 1 to more than 10$^4$ photons of typically 12.4~keV per pixel for an XFEL pulse duration of less than 100 fs, and a radiation tolerance for doses up to 1~GGy (SiO$_2$) for 3 years of operation.

  One question is, if  all charges are collected in the 220~ns between XFEL pulses for the high instantaneous charge-carriers densities, or if pile-up effects appear.
 In~\cite{Becker:2010, Becker:Thesis} the impact of the plasma effect~\cite{Tove:1967}, which occurs for high X-ray densities when the density of electron-hole ($eh$) pairs is large, typically of the order of the doping of the silicon crystal, has been studied.
  From these studies, it has been concluded that for 500~$\upmu $m thick sensors the operating voltage should be at least 500~V in order to assure the complete signal collection in-between XFEL pulses and a sufficient narrow point spread function for the measurement of the shape of narrow Bragg peaks.
 The present work concentrates on the collection of charges produced in the region below the Si-SiO$_2$ interface in segmented  $p^+n$ sensors, where the potential, under certain biasing conditions, has a saddle point and the electric field is zero.
  The multi-channel Transient Current Technique (m-TCT) for charges produced by a  sub-nanosecond laser light of 660~nm (absorption length in silicon of about 3.5~$\upmu $m at room temperature) is used for the studies.

 The same study also allows a detailed investigation of the properties of the accumulation layer, which forms in segmented $p^+n$ sensors at the Si-SiO$_2$ interface, and of the close-by electric field.
  This is the main topic of this manuscript.

 In \cite{Poehlsen:2012} it has been demonstrated that charge carriers produced close to the Si-SiO$_2$ interface can be lost, meaning, that they are not collected by an electrode of the sensor within $\lesssim 100$~ns.
  Depending on the biasing history and on environmental parameters like humidity, situations with losses of electrons, of holes and without losses have been observed.
 The different situations are related to the density of oxide charges, which strongly depends on the X-ray dose with which the sensor has been irradiated, and on the potential distribution on the surface of the sensor's passivation layer, which changes when the biasing voltage is changed.
  Given the high surface resistivity of the passivation layer and its strong dependence on humidity, the time constants for reaching a steady-state of the surface potential can be as long as several days.
 As discussed in detail in \cite{Poehlsen:2012}, the cause of the charge losses is the electric field which, for an electron-accumulation layer points away from the Si-SiO$_2$ interface, and for a hole-accumulation layer towards it.
  In this field  charge carriers drift towards the accumulation layer, and are trapped for times longer than the integration times used in the analysis of the m-TCT data.

 In \cite{Poehlsen1:2012} the time dependence of the electron and hole losses close to Si-SiO$_2$ interface for an non-irradiated  silicon strip sensor and a sensor irradiated by X-rays to 1~MGy (SiO$_2$) have been investigated as function of biasing history and relative humidity.

 In the present work we investigate how many additional charges have to be trapped in the accumulation layers to significantly change the collection of charges from the region close to the Si-SiO$_2$ interface and thus the local electric field, and the time dependence of returning to the steady-state conditions of the accumulation layers.

 The work has been done within the AGIPD collaboration \cite{AGIPD, AGIPD2} which is developing a large-area pixel-detector system for experimentation at the European XFEL and other X-ray sources.

\section{Measurement techniques and analysis}

\subsection{Sensors under investigation}
 \label{sec:sensors}

\begin{figure}
  \centering
	\includegraphics[width=10cm]{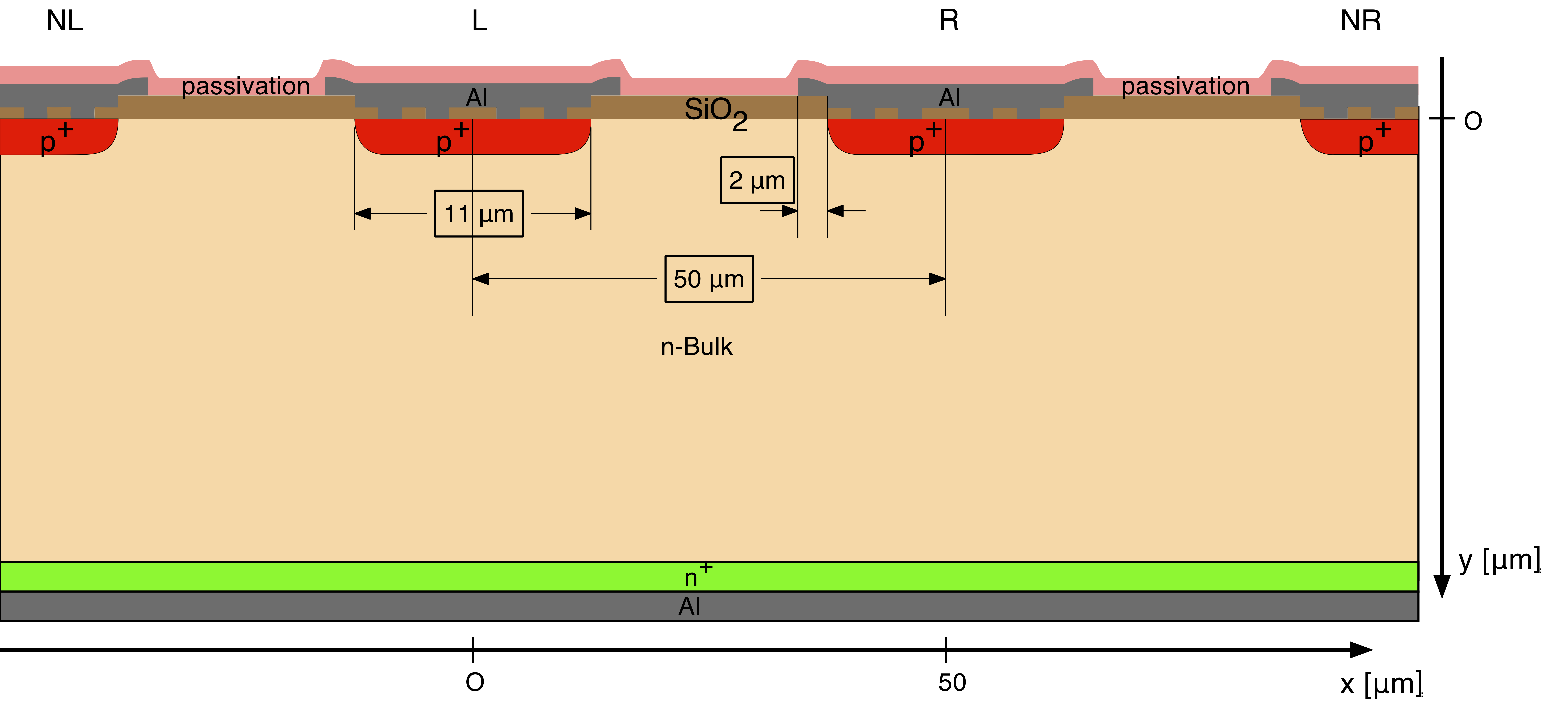}	
     \caption{Schematic cross section of the DC-coupled $p^+n$ sensor, and definition of the $x$ and $y$ coordinates. The drawing is not to scale.}
	\label{fig:sensor3}
 \end{figure}

 The same DC-coupled $p^+n$ strip sensors produced by Hamamatsu \cite{Hamamatsu} as in \cite{Poehlsen:2012, Poehlsen1:2012} were used for the investigations.
  Relevant sensor parameters are listed in Table \ref{tab:sensors3}, and a cross section of the sensor is shown in Figure~\ref{fig:sensor3}.
 The sensors are covered by a passivation layer with openings at the two ends of each strip for bonding.
  One sensor was investigated as produced, and another after irradiation to 1~MGy (SiO$_2$) with 12~keV photons and annealed for 60 minutes at 80$^\circ$C.
 The corresponding values for the oxide-charge density, $N_{ox}$, the integrated interface-trap density, $N_{it}$, and the surface-current density, $J_{surf}$, are listed in Table \ref{tab:irrad3}.
  The values have been derived from measurements on MOS capacitors and gate-controlled diodes fabricated on the same wafer as the sensors \cite{Zhang:2011a, Zhang:2011b, Perrey:Thesis} and scaled to the measurement conditions.

\begin{table}
\centering
  {\renewcommand{\arraystretch}{1.2}
  \renewcommand{\tabcolsep}{0.2cm}
\begin{tabular}{|c|c|}
\hline
sensor parameter 			& value	        \\ \hline \hline
producer 					& Hamamatsu 	\\ \hline
coupling 					& DC 			\\ \hline
pitch 						& 50 $\upmu$m 	\\ \hline
depletion voltage		    & $\sim$~155 V		\\ \hline
doping concentration		& $\sim$~10$^{12}$ cm$^{-3}$	\\ \hline
gap between $p^+$ implants	& 39 $\upmu$m 	\\ \hline
width of $p^+$ implant window	& 11 $\upmu$m	\\ \hline
depth of $p^+$ implant  		& unknown 		\\ \hline
aluminium overhang 			& 2 $\upmu$m	\\ \hline
number of strips 			& 128        	\\ \hline
strip length 				& 7.956 mm 		\\ \hline
sensor thickness			& 450 $\upmu$m	\\ \hline
SiO$_2$ thickness 			& 700 nm     	\\ \hline
passivation layer			& unknown    	\\ \hline
crystal orientation 		& $\langle 1 0 0 \rangle$ \\ \hline
 \end{tabular}}
   \caption{Sensor parameters.}
   \label{tab:sensors3}
\end{table}


\begin {table}
 \centering
  {\renewcommand{\arraystretch}{1.2}
  \renewcommand{\tabcolsep}{0.2cm}
\begin{tabular}{|c|c|c|c|}
 \hline
   X-ray dose  & 0 Gy                          &  1 MGy (60 min. at 80$^\circ $C) \\ \hline \hline
   $N_{ox}$    &  1.3$\cdot$10$^{11}$/cm$^{2}$ & 1.4$\cdot$10$^{12}$/cm$^{2}$ \\ \hline
   $N_{it}$    & 0.87$\cdot$10$^{10}$/cm$^{2}$ & 1.6$\cdot$10$^{12}$/cm$^{2}$ \\ \hline
   $J_{surf}$  & 9.8 nA/cm$^{2}$               & 2.2 $\upmu$A/cm$^{2}$        \\ \hline
 \end{tabular}}
   \caption{Oxide-charge density, $N_{ox}$, interface-trap density integrated over the Si-band gap, $N_{it}$, and surface-current density, $J_{surf}$, obtained from measurements on test structures (a MOS capacitor and a gate-controlled diode). The values for a temperature of 22.9$^\circ$C before and after X-ray irradiation to 1~MGy and annealing for 60 minutes at 80$^\circ$C are presented. The actual measurements were taken at 21.8$^\circ$C and, for the irradiated structures after annealing for 10 minutes at 80$^\circ$C, scaled (scale factor $\sim $~0.7) to above values, which correspond to the measurement conditions of the sensor investigated.}
 \label{tab:irrad3}
\end{table}

\subsection{Experimental setup} \label{sec:setup3}

 To study the charge transport and charge collection in the sensor, the instantaneous currents induced in the electrodes by the moving charges were measured (Transient Current Technique - TCT \cite{Kraner:1993, Becker:2010, Becker:Thesis}).
  The multi-channel TCT setup, described in detail in \cite{Becker:Thesis}, has been used for the measurements.
 The bias voltage was applied on the $n^+$ rear contact of the sensor.
  The current signal was read out on the rear contact and on 2 strips on the front side using Agilent 8496G attenuators, Femto HSA-X-2-40 current amplifiers and a Tektronix digital oscilloscope with $2.5$ GHz bandwidth (DPO 7254).
 The readout strips were grounded through the DC-coupled amplifiers ($\sim$~50~$\Omega$ input impedance).
  The seven strips to the right and the seven strips to the left of the strips read out were connected to ground by 50~$\Omega$ resistors.

 \begin{figure}[b]
  \centering
	\includegraphics[width=12cm]{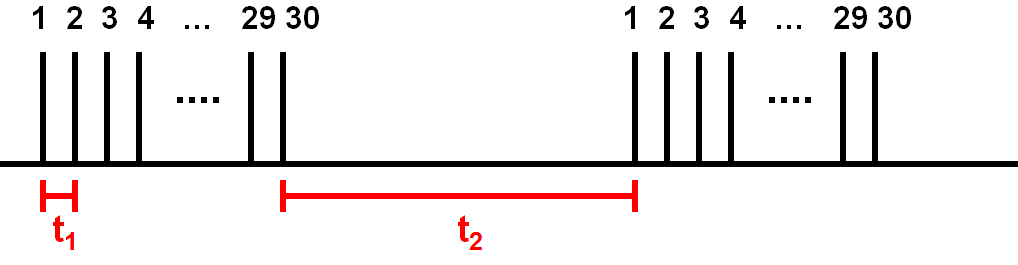}	
     \caption{Schematic of the pulse structure. The laser was operated in burst mode with 30 pulses per burst. Pulses inside a burst were separated by the time interval $t_1$, and the time interval between bursts was $t_2$.}
	\label{fig:burst}
 \end{figure}

  Electron-hole ($eh$) pairs were generated in the sensor close to its surface in-between the readout strips by red light from a laser focussed to an $rms$ of 3~$\upmu$m.
 The wavelength of the light was 660~nm, which has an absorption length in silicon at room temperature of approximately 3.5~$\upmu$m.
  We note that 1~keV X-rays have a similar absorption length in silicon.
  The number of generated $eh$~pairs was controlled by optical filters.
 For most of the measurements presented approximately 130~000 $eh$~pairs were generated, corresponding to 470~X-rays of 1~keV.
 The laser was used in burst mode with 30 pulses per burst.
  The pulse structure is shown in Figure \ref{fig:burst}.
 The pulses in a burst were separated by $t_1 = 50$~ns, and the time interval between bursts $t_2 \approx 1$~ms, if not stated otherwise.
  To study the time dependence of the return to steady-state conditions after charges have been trapped, $t_1$ or $t_2$ were varied, with the other parameters fixed.
 For longer recovery times $t_2$ was varied between 500~ns and 10~ms and the signal from the first pulse of the burst was analysed.
The recovery time $\Delta t$ is defined as the time interval between the pulse analysed and its preceding pulse.
 Hence $\Delta t = t_2$ if the first pulse is analysed.
  For short recovery times $t_2$ was set to 1~ms, $t_1$ varied between 50 and 500~ns, and the signal from pulse 30 analysed. In this case we have $\Delta t = t_1$.
 In this way two different measurements are available for $\Delta t = 500$~ns.

 \begin{figure}
  \centering
	\includegraphics[width=14cm]{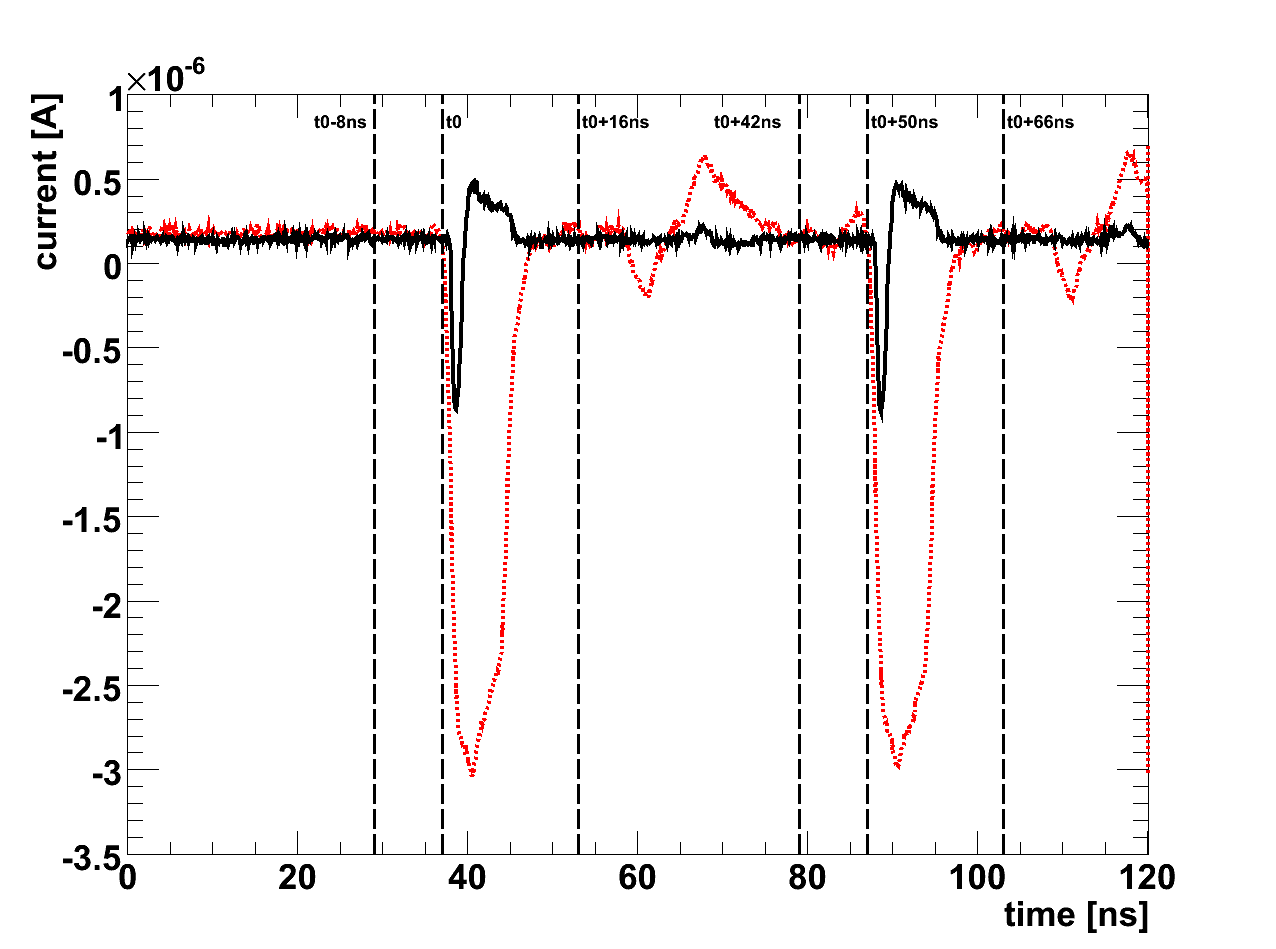}	
    \caption{Current transients for the first two pulses of the burst for strip $L$ (black solid line) and the rear contact (red dotted line) for 130~000~$eh$~pairs generated at $x_0 = 75$~$ \upmu$m for $t_1 = 50$~ns and $t_2 = 1$~ms. The vertical lines indicate the limits used to determine the base line and the signal. The results shown are for the non-irradiated sensor in conditions "dried~@~500~V" biased to 200~V.}
	\label{fig:pulse1}
 \end{figure}

 \subsection{Analysis method} \label{sec:analysis}

 Figure~\ref{fig:pulse1} shows for the non-irradiated sensor biased to 200~V the current transients of the first two pulses of a pulse train  measured at strip $L$ and at the rear contact for 130~000~$eh$~pairs generated at $x_0 = 75$~$ \upmu$m, half way between the readout strips $R$ and $NR$, as shown in Figure~\ref{fig:sensor3}.
  The red dotted line in Figure~\ref{fig:pulse1} shows the signal measured at the rear contact, and the black solid line the signal from strip $L$ at $x = 0$, which is 1.5~times the strip pitch away from the position where the $eh$~pairs were generated.

  The signals are the sums of the currents induced by the holes, which drift to the $p^+$ strips, and the electrons, which drift to the $n^+$-rear contact.
 The holes are collected quickly, because the distance between the readout strips and the place where they were generated is small.
  The electron signals are significantly longer, as electrons have to traverse the entire sensor to reach the rear contact.
 The current transient on strip $L$ is the sum of a short negative signal from the holes drifting to strips $R$ and $NR$ and a slower positive signal from the electrons drifting to the rear contact.
  In the rear contact the holes as well as the electrons induce negative signals.
  The bi-polar signals, starting approximately 20~ns after the start of the signal pulse, are due to reflections from the amplifiers, which were connected to the electrodes by 2~m long cables.
 The noise and the reflection for the rear contact are significantly higher than for the signal from the readout strips.
  This is due to the higher capacitance of the rear contact and the bias-T used to decouple the high voltage.

In the analysis the induced charge for the i-th pulse in a burst, $Q_i$, is calculated off-line by integrating the current over the time interval $\delta t$ and subtracting the baseline current:
  \begin{equation}
   Q_i = \int^{\tau_i + \delta t}_{\tau_i}  (I - I_{baseline}) \cdot \mathrm{dt} 
    \quad \text{with} \quad 
   I_{baseline} = \frac{\int_{\tau_i - 8\,\text{ns}}^{\tau_i} I \cdot \mathrm{dt}}{8\,\text{ns}},
    \quad \text{and} \quad 
   \tau_i = t_0 + (i-1) \cdot t_1.
   \label{Q}
  \end{equation}
As indicated in Figure~\ref{fig:pulse1}, $t_0$ is the time shortly before the first pulse starts and a value $\delta t = 16$~ns was chosen for the measurements with 130~000 $eh$~pairs.
For the measurements in which the number of $eh$~pairs was varied between $10^5$ and $10^7$, $\delta t = 40$~ns had to be chosen in order to collect the entire charge.

 The number of charge carriers lost is obtained from $Q^L$, the charge induced in strip $L$ in the following way:
  The integral of the hole signal is $Q_h^L = N_h \cdot q_0 \cdot (0 - \Phi _w^L(x_0))$.
 $N_h$ is the number of holes collected, $q_0$ the elementary charge, and the term in parenthesis the difference of the weighting potential $\Phi _w^L$ for the readout strip $L$ at the strips $R$ and $NR$ where the holes are collected ($\Phi _w^L(R) = \Phi _w^L(NR) = 0$), and $\Phi _w^L(x_0)$, the weighting potential at the position $x_0$, where the holes were generated.
  The charge induced by the electrons is $Q_e^L = N_e \cdot (-q_0) \cdot (0 - \Phi _w^L(x_0))$, where $N_e$ is the number of electrons collected at the rear contact.
 The total charge induced on strip $L$  is:
 \begin{equation}
   Q^L = Q_e^L + Q_h^L = (N_e - N_h) \cdot q_0 \cdot \Phi _w^L(x_0).
  \label{Q_sum}
 \end{equation}
 If all holes and electrons are collected $N_e = N_h$ and $Q^L = 0$.
  For incomplete charge collection, assuming that there is negligible $eh$ recombination and only electrons or only holes are lost, the amount of charge lost is given by:
 \begin{equation}
  Q_{lost} = (N_e - N_h) \cdot q_0 = Q^L/\Phi _w^L(x_0).
  \label{Q_lost}
 \end{equation}
 If more electrons than holes are collected $N_e > N_h$, $Q_{lost}$ is positive, and the number of holes lost is obtained from $N_h^{lost} = Q_{lost} / q_0$.
  In a similar way, for $N_e < N_h$ the number of electrons lost is  $N_e^{lost} = -Q_{lost} / q_0$.

 In this paper measurements from strip $L$ for light injected at the positions $x_0 = 40$~$\upmu$m and $x_0 = 75$~$\upmu$m are presented.
  For the corresponding weighting potentials $\Phi _w^L(x_0)$ values of 0.35 and 0.05 are used.
 For the analysis neither the signals from the rear contact nor from strip $R$ are used, but it has been verified that the corresponding signals agree with the results of the analysis from strip $L$.
  For more details on this method of determining the charge losses and on the way the values of the weighting potentials were obtained, we refer to \cite{Poehlsen:2012}.

 \section{Results}
  \label{chapter:results}

 First, the three biasing and environmental conditions under which the measurements have been performed, are defined.
  Then, for the non-irradiated and for the irradiated (1~MGy) sensor in the three experimental conditions it is shown how the charge losses for 130~000 $eh$~pairs generated per pulse depend on the pulse number in the burst.
 Next, the time dependence of the recovery of the charge losses to the situation for the first pulse of the pulse train is investigated.
  Finally, for the irradiated sensor, the dependence of the charge losses on pulse number as function of the number of generated charge carriers in the range between $10^5$ and $10^7$ is shown.
 A discussion and qualitative explanations of the results are found in Section~\ref{chapter:discussion}.

  \subsection{Measurement conditions}
   \label{sec:conditions}

 As discussed in detail in~\cite{Poehlsen:2012, Poehlsen1:2012}, the observed charge losses depend on the X-ray-radiation damage and on the charge distributions inside and on top of the passivation layer.
  The latter changes when the biasing voltage is changed.
 After changing the biasing voltage, steady-state conditions are reached on top of the passivation layer after a time interval which, due to the dependence of the surface resistivity on humidity, strongly depends on the ambient relative humidity.
  In a dry atmosphere or in vacuum, this time can be as long as several days, whereas in a humid atmosphere, it can be as short as minutes.
 All measurements were performed at 200~V.
  The same biasing and environmental conditions were already used in~\cite{Poehlsen:2012}:

 \begin{itemize}
   \item "humid":~Sensor biased to 200~V and kept in a humid atmosphere for $>$~2~hours (relative humidity $>$~60~$\%$), i.e. in steady-state conditions on top of the passivation layer,
    \item "dried~@~0~V":~Sensor stored at 0~V for a long time to reach steady-state conditions at 0~V, then kept in a dry atmosphere for $>$~1~hour (relative humidity $<$~5~$\%$), and then biased to 200~V for the measurements; thus the charge distribution on top of the passivation layer corresponds to the 0~V condition,
   \item "dried~@~500~V":~Sensor kept for $>$~2~hours at 500~V in a humid atmosphere (relative humidity $>$~60~$\%$) to reach steady-state conditions, then dried for $>$~1~hour, and afterwards biased at 200~V in a dry atmosphere for the measurements.
 \end{itemize}

 \begin{figure}
  \centering
   \includegraphics[width=8cm]{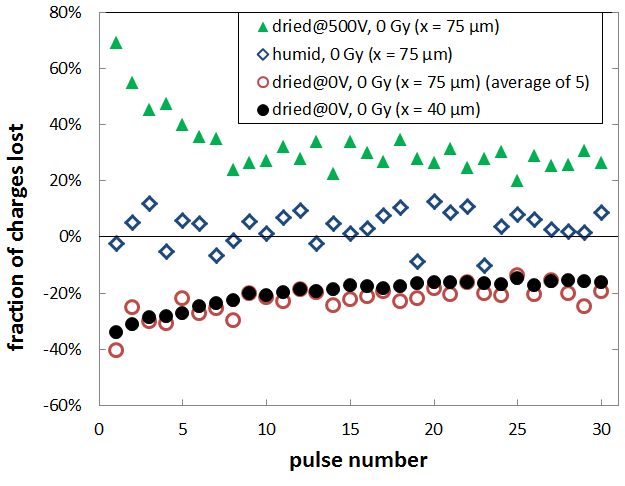}	
    \caption{Fraction of charges lost as a function of pulse number in the burst for the non-irradiated sensor for $\sim$~130~000~$eh$ pairs generated per pulse at $x_0 = 75$~$\upmu$m. In addition, some measurements with the laser at $x_0 = 40$~$\upmu$m are shown. The sensor was biased at 200~V. The nomenclature characterizing the different measurement conditions are explained in the text. Positive values correspond to hole losses and negative to electron losses.}
   \label{fig:0Gy_pulse}
 \end{figure}

 \begin{figure}
  \centering
	\includegraphics[width=8cm]{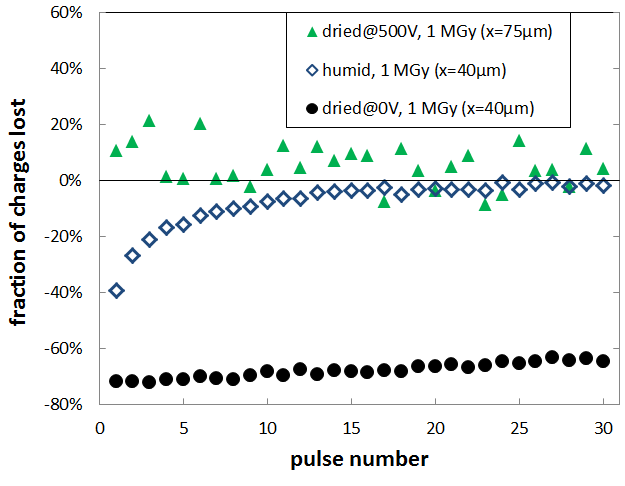}	
     \caption{Fraction of charges lost as a function of pulse number for the irradiated sensor (1MGy). The other conditions are the same as for Figure~\ref{fig:0Gy_pulse}.}
  \label{fig:1MGy_pulse}
 \end{figure}

   \subsection{Charge losses as function of pulse number}
     \label{sec:pulse_number}

 Figures~\ref{fig:0Gy_pulse} and \ref{fig:1MGy_pulse} show for the non-irradiated and the irradiated sensor for the three experimental conditions and $\sim$~130~000~$eh$ pairs generated per pulse, the fraction of charges lost as function of the pulse number in the burst. The main results are summarised in Table~\ref{tab:pulse_number}.
   The parameters of the burst mode were a time between the pulses $t_1=50$~ns, and a time between the bursts  $t_2=1$~ms.
  As will be shown later the value $t_2=1$~ms is sufficient that the charge losses have recovered to the steady-state values before the first pulse of the following burst.

  In Figure~\ref{fig:0Gy_pulse} the fractions of charges lost for the two laser positions, $x_0 = 40$~$\upmu $m and $x_0 = 75$~$\upmu $m, for the condition "dried~@~0~V" are shown.
 It can be seen, that the fluctuations for $x_0 = 40$~$\upmu $m are much smaller than for $x_0 = 75$~$\upmu $m.
  The reason is the difference in weighting potential, which is in the denominator in Equation~(\ref{Q_lost}). It is 0.35 for $x_0 = 40$~$\upmu $m and 0.05 for $x_0 = 75$~$\upmu $m.
   However, $x_0 = 40$~$\upmu $m is only 15~$\upmu $m away from the center between the strips $R$ and $L$, and not for all conditions it can be assured, that no holes reach the readout strip $L$ by diffusion, which is assumed in the analysis.
 In \cite{Poehlsen:2012} it has been shown that there are situations where the diffusion of the holes is sufficiently small, so that the measurements at $x_0 = 40$~$\upmu $m give reliable results for the charge losses.
  This is the case for "dried~@~0~V", and the results are compatible with the measurements at $x_0 = 75$~$\upmu $m.
 In the following, if no holes diffuse to strip $L$ the results for $x_0 = 40$~$\upmu $m are shown, else the results for $x_0 = 75$~$\upmu $m.


\begin {table}
 \centering
  {\renewcommand{\arraystretch}{1.2}
  \renewcommand{\tabcolsep}{0.2cm}
\begin{tabular}{|c|c|c|c|c|c|c|}
  \hline
           &carrier& pulse 1 & saturation & pulse no.  & no.  charges lost\\
 condition & type   & losses & losses & saturation & for saturation  \\ \hline \hline
0 Gy, dried@500 V &$h$ & 70 \%   & 25 \%   & $\sim $ 8  & $\sim$ 500 000   \\ \hline
0 Gy, humid       &$-$ & $<$ 10 \% & $<$ 10 \% &$-$ &$-$        \\ \hline
0 Gy, dried@0 V   &$e$ & 35 \%   & 20 \%   & $\sim $  8  & $\sim$ 300 000   \\ \hline \hline
1 MGy, dried@500 V&$-$ & $<$ 20 \%&$<$ 20 \% &$-$  &$-$       \\ \hline
1 MGy, humid      &$e$ & 40 \%   & $<$ 5 \% & $\sim $  15 & $\sim$ 250 000   \\ \hline
1 MGy, dried@0 V  &$e$ & 70 \%    & -     & $>$ 30&$>$ 2 500 000 \\ \hline
 \end{tabular}}
   \caption{Summary on the dependence of the charge losses on pulse number for 130~000 $eh$~pairs produced, extracted from Figures~\ref{fig:0Gy_pulse} and \ref{fig:1MGy_pulse}. Presented are, for the non-irradiated and the irradiated sensor and three measurement conditions, the type of charge carriers lost, the initial losses for pulse number one, the saturation value of the charge losses, the pulse number at which the losses reach saturation, and the total number of charges lost until the saturation is reached.}
 \label{tab:pulse_number}
\end{table}

 \begin{figure}[b]
  \centering
   \includegraphics[width=15cm]{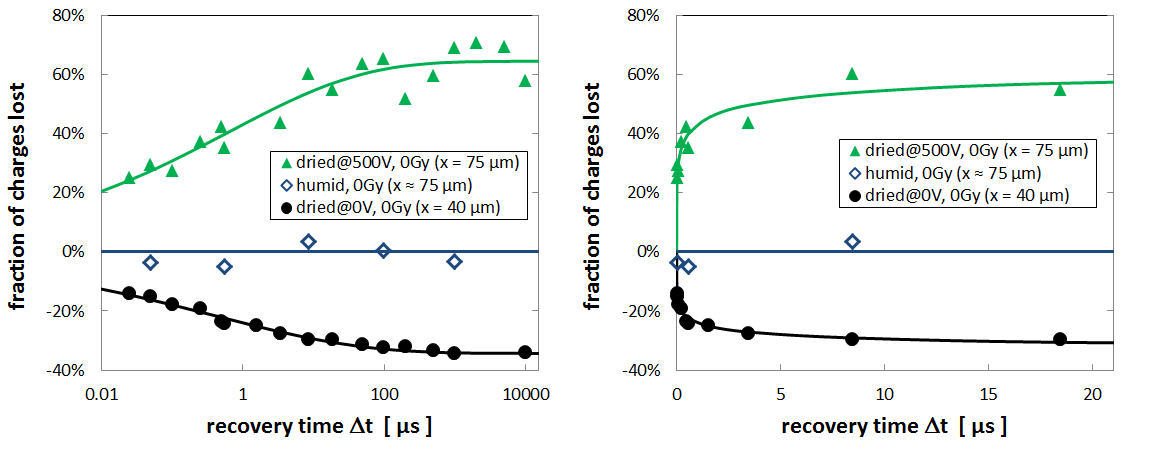}
    \caption{Fraction of charges lost as a function of the recovery time $\Delta t$ for the non-irradiated sensor biased at 200~V for $\sim$~130~000 $eh$~pairs generated. Left: Logarithmic time axis. Right:  Linear time axis.}
   \label{fig:0Gy_time}
 \end{figure}

   \subsection{Charge losses as function of recovery time}
     \label{sec:pulse_intensity}

 \begin{figure}
  \centering
   \includegraphics[width=15cm]{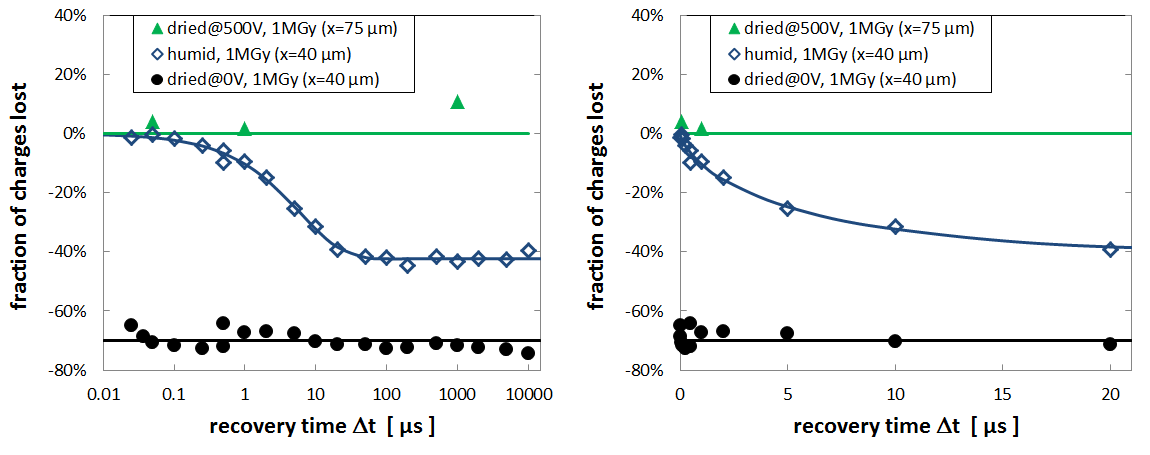}
     \caption{Fraction of charges lost as a function of the recovery time $\Delta t$ for the irradiated sensor biased at 200~V for $\sim$~130~000 $eh$~pairs generated. Left: Logarithmic time axis. Right:  Linear time axis.}
  \label{fig:1MGy_time}
 \end{figure}

 Figures~\ref{fig:0Gy_time} and \ref{fig:1MGy_time} show the fraction of  charges lost as a function of the recovery time $\Delta t$, defined in Section~\ref{sec:setup3}, for the irradiated and non-irradiated sensor biased at 200~V and $\sim$~130~000 $eh$~pairs generated.
  For the measurements at $\Delta t = 500$~ns there are two data points.
 As discussed in Section~\ref{sec:setup3}, one is obtained from pulse number 30 for the laser timing $t_1 = 500$~ns and $t_2 = 1$~ms, the other from pulse number 1 for the timing $t_1 = 50$~ns and $t_2 = 500$~ns.
  It is seen that the values are compatible.
 Smooth transitions from the reduced charge losses at short recovery times to the larger steady-state losses, corresponding to the losses for the first pulse in Figures~\ref{fig:0Gy_pulse} and \ref{fig:1MGy_pulse}, are observed.

 In order to obtain a quantitative description of the measurements, they are fitted by the phenomenological function
  \begin{equation}
	f_{lost}(\Delta t)  = f_{lost}^\infty \left(1-e^{-(\Delta t/t_0)^{p}}\right),
   \label{N_lost_fit}
  \end{equation}
 with the free parameters, the steady-state fraction of charges lost, $f_{lost}^\infty$, the time constant, $t_0$, and the power in the exponent, $p$.
  The fit results are presented in Table~\ref{tab:recovery}.
 The discussion of the results is  postponed to Section~\ref{chapter:discussion}.

\begin {table}
 \centering
 {\renewcommand{\arraystretch}{1.4}
  \renewcommand{\tabcolsep}{0.2cm}
   \begin{tabular}{|c|c|c|c|c|}
   \hline
 condition & type &  $f_{lost}^\infty [\%]$ & $t_0$ [$\upmu $s]& $p$ \\ \hline \hline
0 Gy, dried@500 V &$h$ &$ 64.2 \pm 1.8$   & $0.65_{-0.26}^{+0.57}$	& $0.23 \pm 0.06$ \\ \hline
0 Gy, humid       &$-$ & $<$ 10          & $-$           					& $-$                     \\ \hline
0 Gy, dried@0 V   &$e$ & $ -34.3 \pm 0.3$ &$ 0.42_{-0.05}^{+0.06}$	&$0.21 \pm 0.013$ \\ \hline \hline
1 MGy, dried@500 V&$-$ & $<$ 20          & $-$          						& $-$                      \\ \hline
1 MGy, humid      &$e$ & $ - 45.6 \pm 0.5$& $6.0 \pm 0.4$ 					& $0.71^{+0.05}_{-0.03}$  \\ \hline
1 MGy, dried@0 V  &$e$ & $\sim$~$73$			& $-$ 										& $-$                   \\ \hline
 \end{tabular}}
   \caption{Summary of the dependence of the charge losses on recovery time $\Delta t$ for 130~000 $eh$~pairs produced, extracted from Figures~\ref{fig:0Gy_time} and \ref{fig:1MGy_time}. Presented are, for the irradiated and non-irradiated sensor and three measurement conditions, the type of charge carriers lost, and the parameters obtained when fitting Equation~(\ref{N_lost_fit}) to the data; the steady-state fraction of charges lost, $f_{lost}^\infty$, the time constant, $t_0$, and the power in the exponent, $p$. In three cases no or little dependence on recovery time is found and the data described by a constant.}
 \label{tab:recovery}
\end{table}

   \subsection{Effects of high charge densities} \label{sec:high_density}

\begin{figure}[b]
  \centering
	\includegraphics[width=7.3cm]{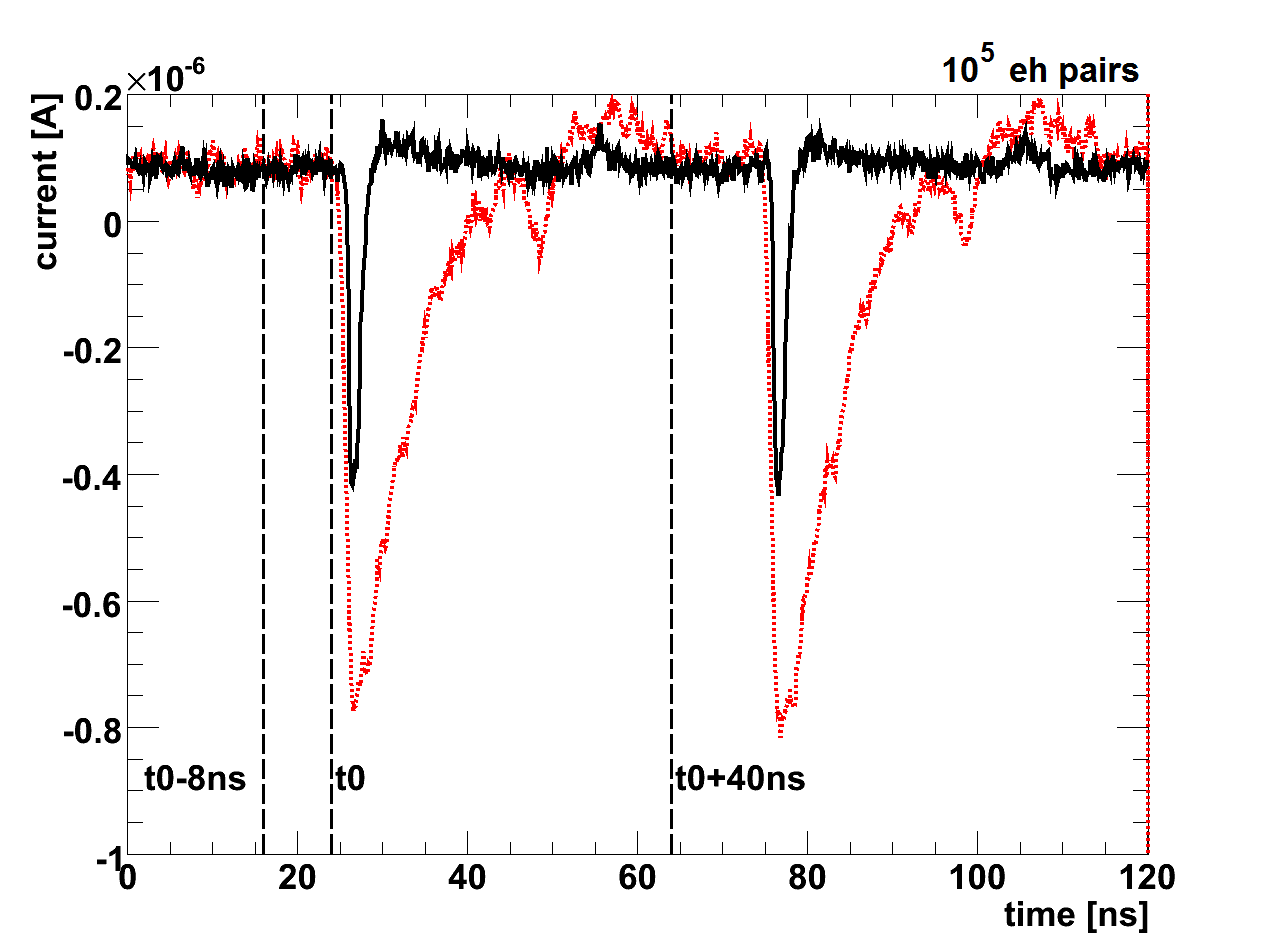}	
	\includegraphics[width=7.3cm]{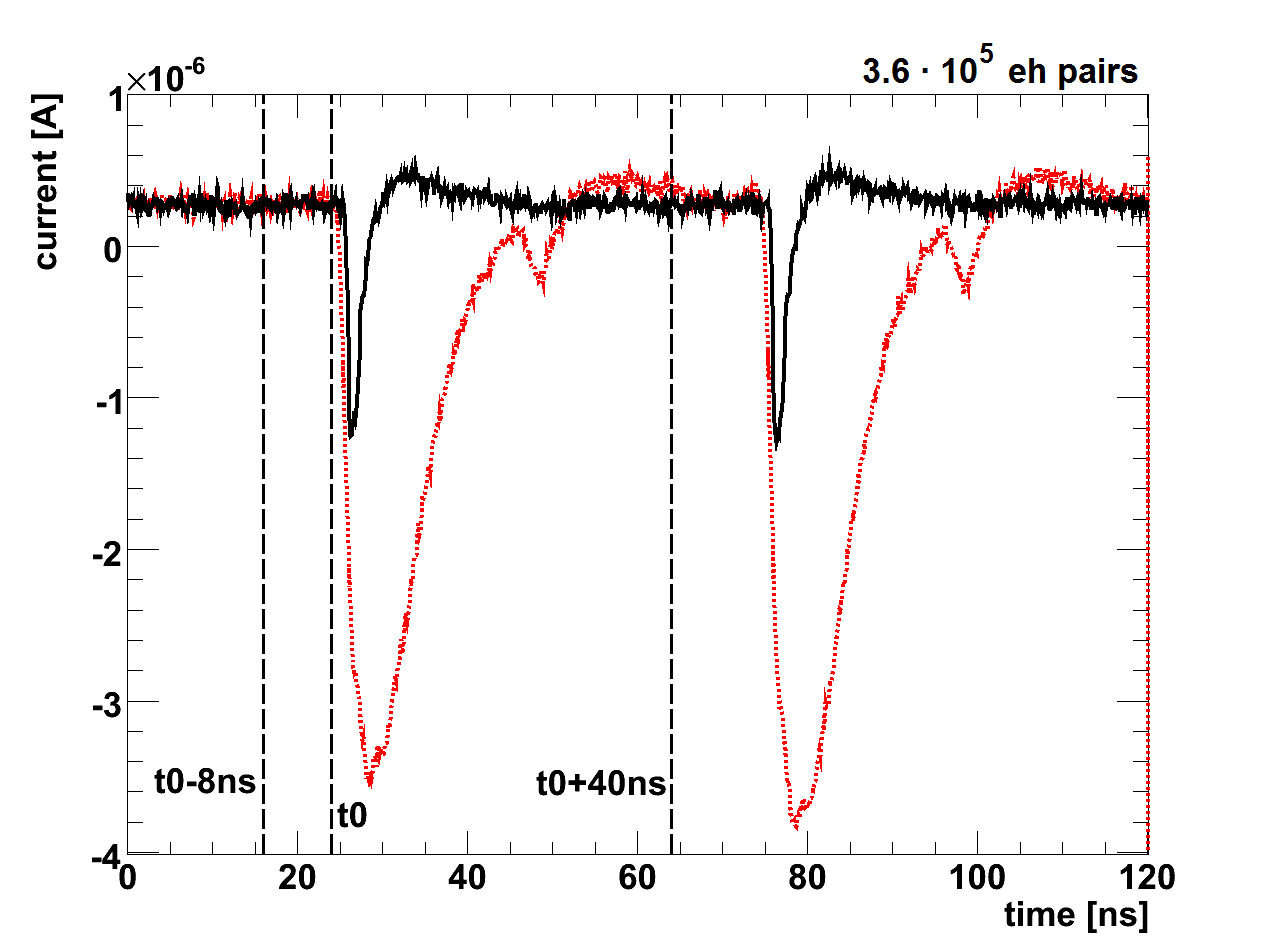}
	\includegraphics[width=7.3cm]{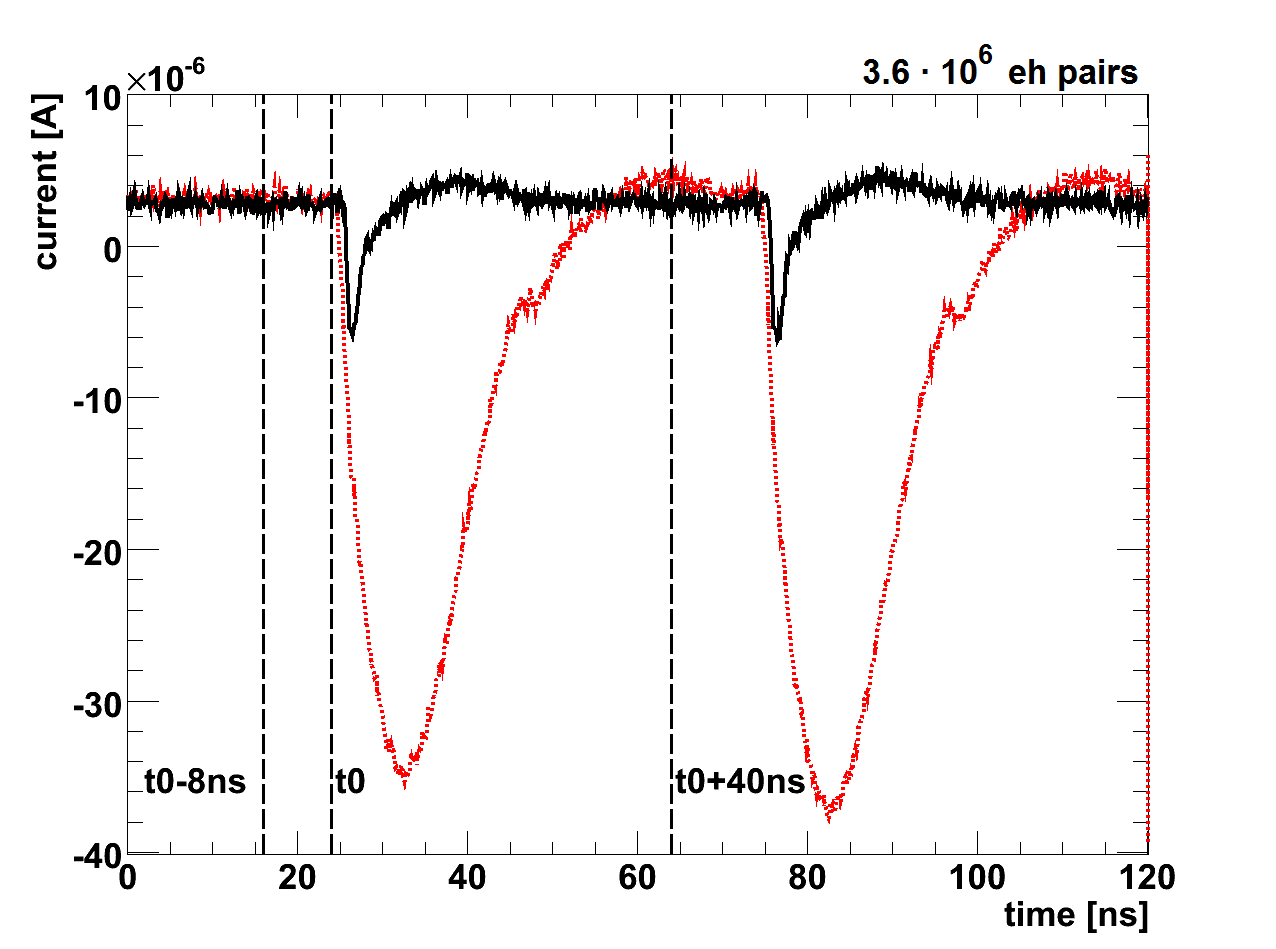}
	\includegraphics[width=7.3cm]{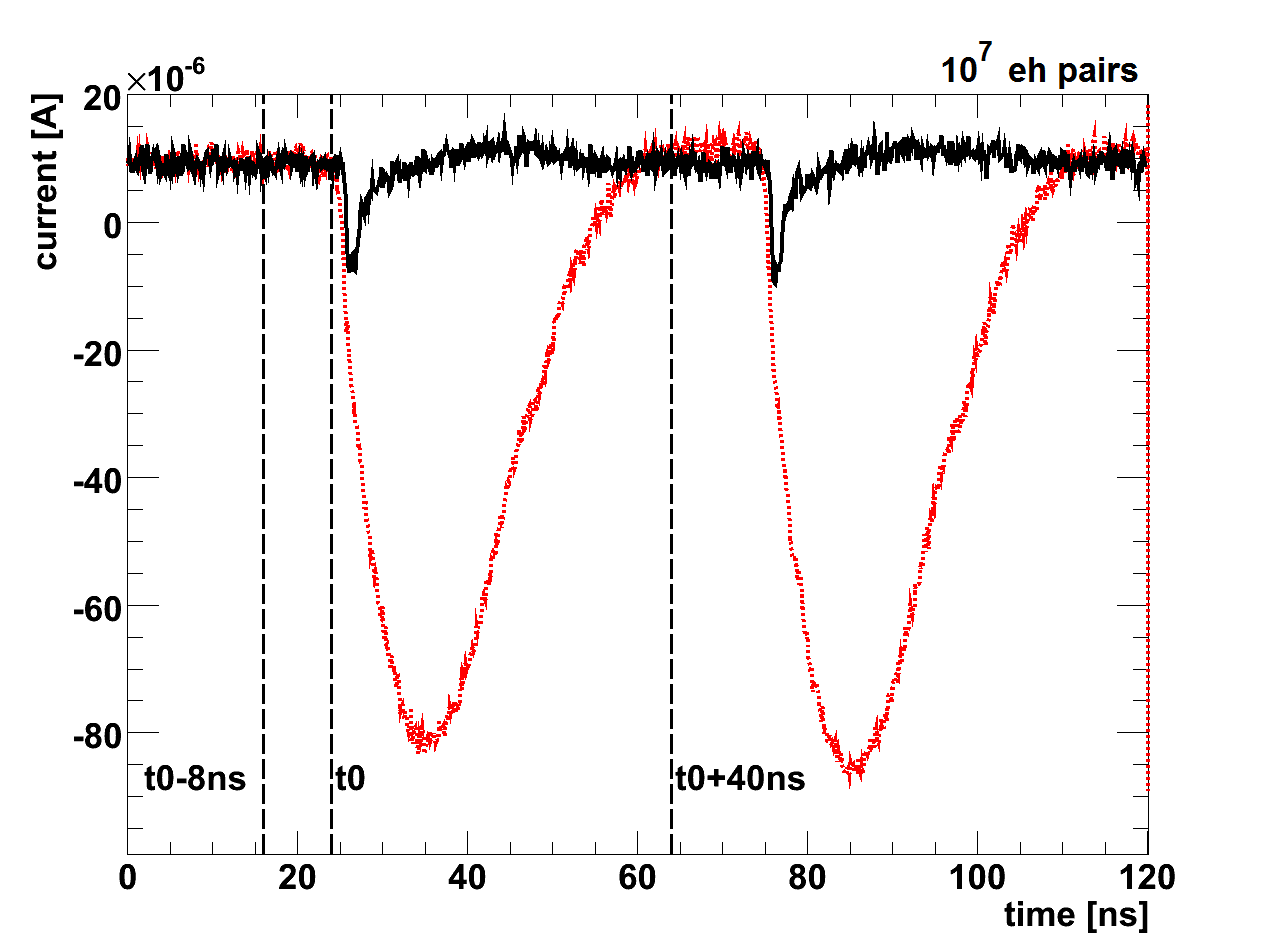}
     \caption{Current transients for the first two pulses of the pulse train for strip $L$ (black solid line) and the rear contact (red dotted line) as function of the number of $eh$~pairs produced at $x_0 = 75$~$ \upmu$m for $t_1 = 50$~ns and $t_2 = 1$~ms. The vertical lines indicate the limits used to determine the base line ($t_0 - 8$~ns to $t_0$) and the signal ($t_0$ to $t_0 + 40$~ns). The measurements were made with the irradiated sensor in the condition "dried~@~0~V" biased to 200~V.
     Top left: $10^5$~$eh$~pairs.
     Top right: $3.6 \cdot 10^5$~$eh$~pairs.
     Bottom left: $3.6 \cdot 10^6$~$eh$~pairs.
     Bottom right: $10^7$~$eh$~pairs.}
	\label{fig:1Million}
 \end{figure}

 For the study of one consequence of the plasma effect, the increase of the pulse length, Figure~\ref{fig:1Million} shows the current transients of the first two pulses of the pulse train for the readout strip $L$ and the rear contact, for $10^5$, $3.6 \cdot 10^5$, $3.6 \cdot 10^6$ and $10^7$~$eh$~pairs produced at $x_0 = 75$~$ \upmu$m for the irradiated sensor biased to 200~V in the condition "dried~@~0~V". We note that the condition "dried~@~0~V" corresponds to operation conditions typical for sensors.

 Whereas the shapes of the signals from the rear contact (red dotted lines), which are mainly due to the electrons, are similar for $10^5$ and $3.6 \cdot 10^5$~$eh$~pairs, a significant change is observed for higher intensities.
  The signal peaks at $\sim 10$~ns, compared to $\sim 2$~ns, and the signal extends up to $\sim 35$~ns compared to $\lesssim 20$~ns.
 Also the signals from strip $L$ (black solid lines) change significantly. The short negative signals due to the holes moving to strips $R$ and $NR$ and the slower positive signals are very much reduced when normalised to the number of $eh$~pairs generated.
  The reason is that both electrons and holes are trapped in the $eh$~plasma, which dissolves by ambipolar diffusion, and the positive electron signal is to a good extent compensated by the negative signal induced by the holes moving towards strips $R$ and $NR$.

 Next we have investigated for the irradiated sensor in the condition "dried~@~0~V", where electron losses of $\sim 70$~\% with little dependence on pulse number and recovery time had been observed, how the number of generated $eh$~pairs influences the charge losses as function of pulse number.
  The laser was used in burst mode with 30~pulses with the parameters $t_1 = 50$~ns and $t_2 = 1$~ms, and the number of $eh$~pairs generated at $x_0 = 40$~$\upmu$m was varied between $10^5$ and $10^7$.
 In order to take into account the increase of the pulse length due to the plasma effect, for this analysis the integration time $\delta t$ in Equation~(\ref{Q}) was increased to 40~ns, as indicated in Figure~\ref{fig:1Million}.

 \begin{figure}
  \centering
   \includegraphics[width=8cm]{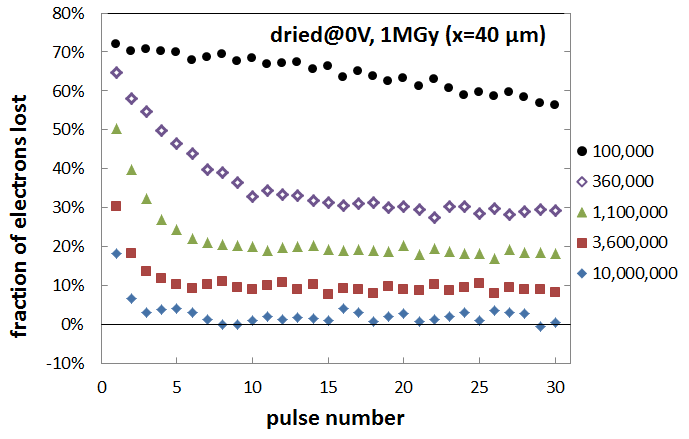}	
    \caption{Fraction of electrons lost as function of the pulse number in the burst for the irradiated sensor biased at 200~V for the condition "dried at 0~V". Between $10^5$ and $10^7$~$eh$~pairs per pulse were generated at $x_0 = 40$~$\upmu$m with pulse spacing $t_1=50$~ns and burst spacing $t_2=1$~ms.}
  \label{fig:1MGy_intensity}
 \end{figure}

 Figure~\ref{fig:1MGy_intensity} shows the fraction of electrons lost as function of the pulse number for different numbers of $eh$~pairs generated per pulse.
  For $10^5$~$eh$~pairs the number of electrons lost per pulse decreases with pulse number from $\sim$~70~\% to $\sim$~60~\% without reaching a constant value up to 30~pulses. This is similar to the data presented in Figure~\ref{fig:1MGy_pulse}.
 For higher numbers of generated $eh$~pairs, the fraction of electrons lost for the first pulse decreases, and a strong further decrease is observed for the following pulses.
  The values at high pulse numbers also decrease with the numbers of generated $eh$~pairs. For $3.6 \cdot 10^5$ a saturation value of $\sim$~30~\% is obtained. For $10^7$ it is as small as $\sim$~2~\%.
 The observation that the fraction of electrons lost decreases with increasing number of generated $eh$~pairs already for the first pulse agrees with the expectation, that the local electric field changes already before the charges from that particular pulse are collected.

\section{Discussion}
 \label{chapter:discussion}

   \subsection{Plasma effect} \label{sec:discussion_plasma}


 With respect to the questions, if  there are pile-up effects due to charges trapped in the region below the Si-SiO$_2$ interface, we conclude from Figure~\ref{fig:1Million}, that a significant lengthening of the current pulse occurs only when more than $3.6 \cdot 10^5$~$eh$~pairs are produced by the laser.
  For the AGIPD sensor the X-rays enter through the $n^+$ rear contact, and only $\sim$~0.3~\% of 12.4~keV X-rays (absorption length $\sim $~250~$\upmu $m in silicon) interact in the $\sim $~5~$\upmu $m close to the Si-SiO$_2$ interface, and thus $\sim $~$3.6 \cdot 10^4$ 12.4~keV X-rays are required to produce $3.6 \cdot 10^5$~$eh$~pairs there.
 We conclude, that the low-field region close to the Si-SiO$_2$ interface does not result in increased pulse lengths due to the plasma effect for the situation expected at the European XFEL, and that the conclusions of \cite{Becker:2010} remain valid.


 \begin{figure}
  \centering
	\includegraphics[width=7.3cm]{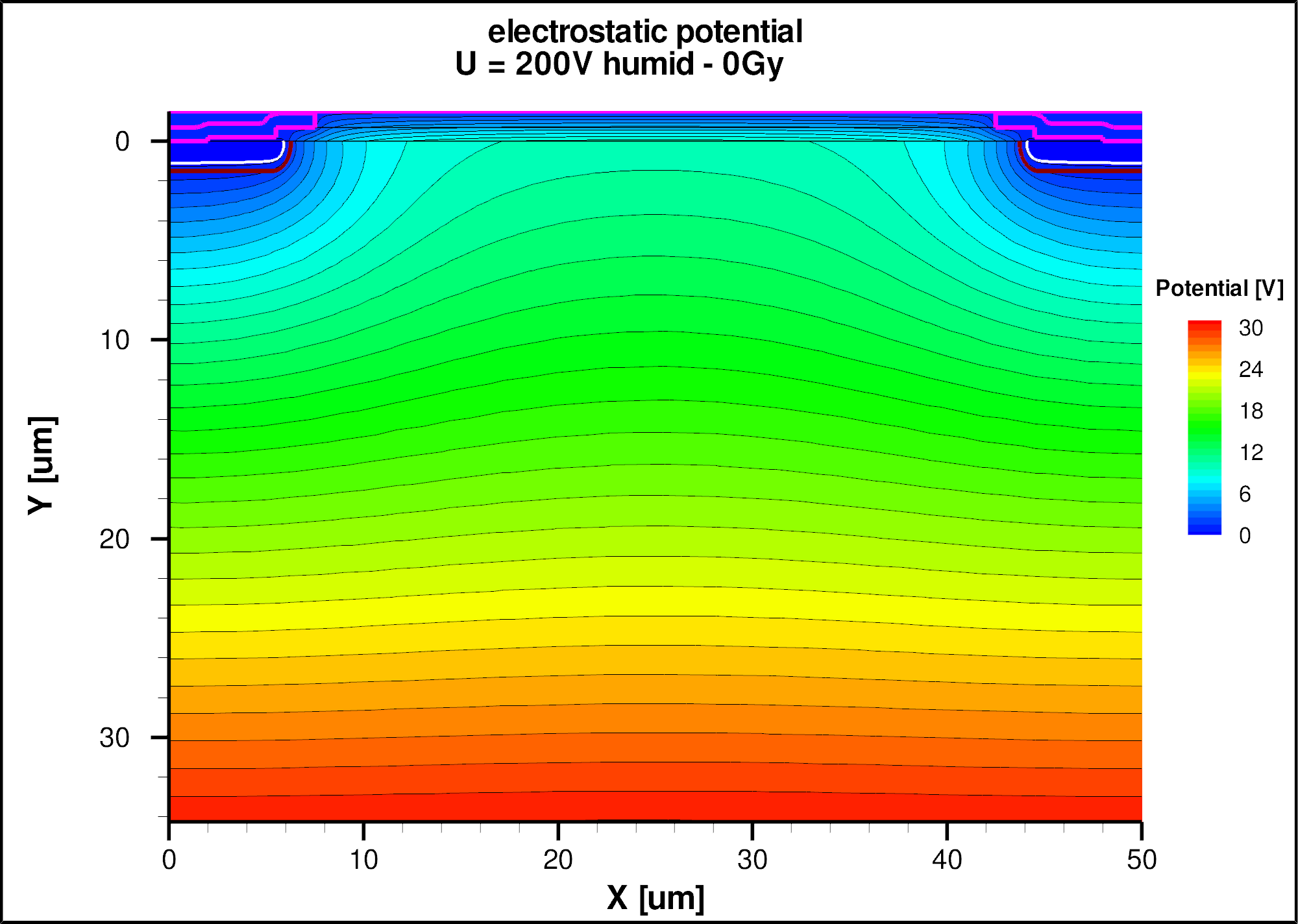}
   \caption{Simulated potential distribution for the non-irradiated sensor biased to 200~V with the biasing condition "humid". See Figure \ref{fig:sensor3} for the coordinate system.}
  \label{fig:potential}
 \end{figure}

 \begin{figure}[b]
  \centering
   \includegraphics[width=7.3cm]{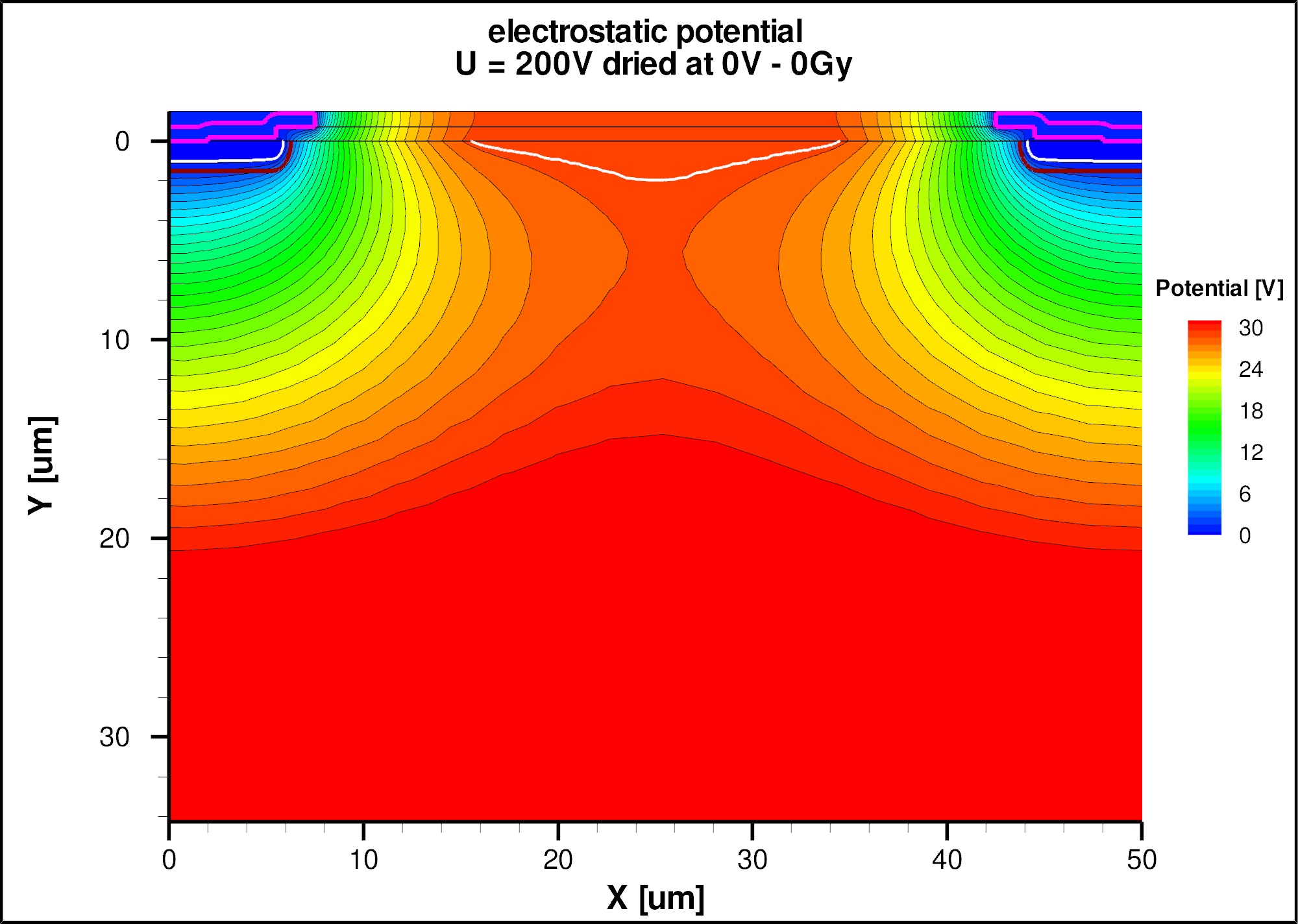}
   \includegraphics[width=7.3cm]{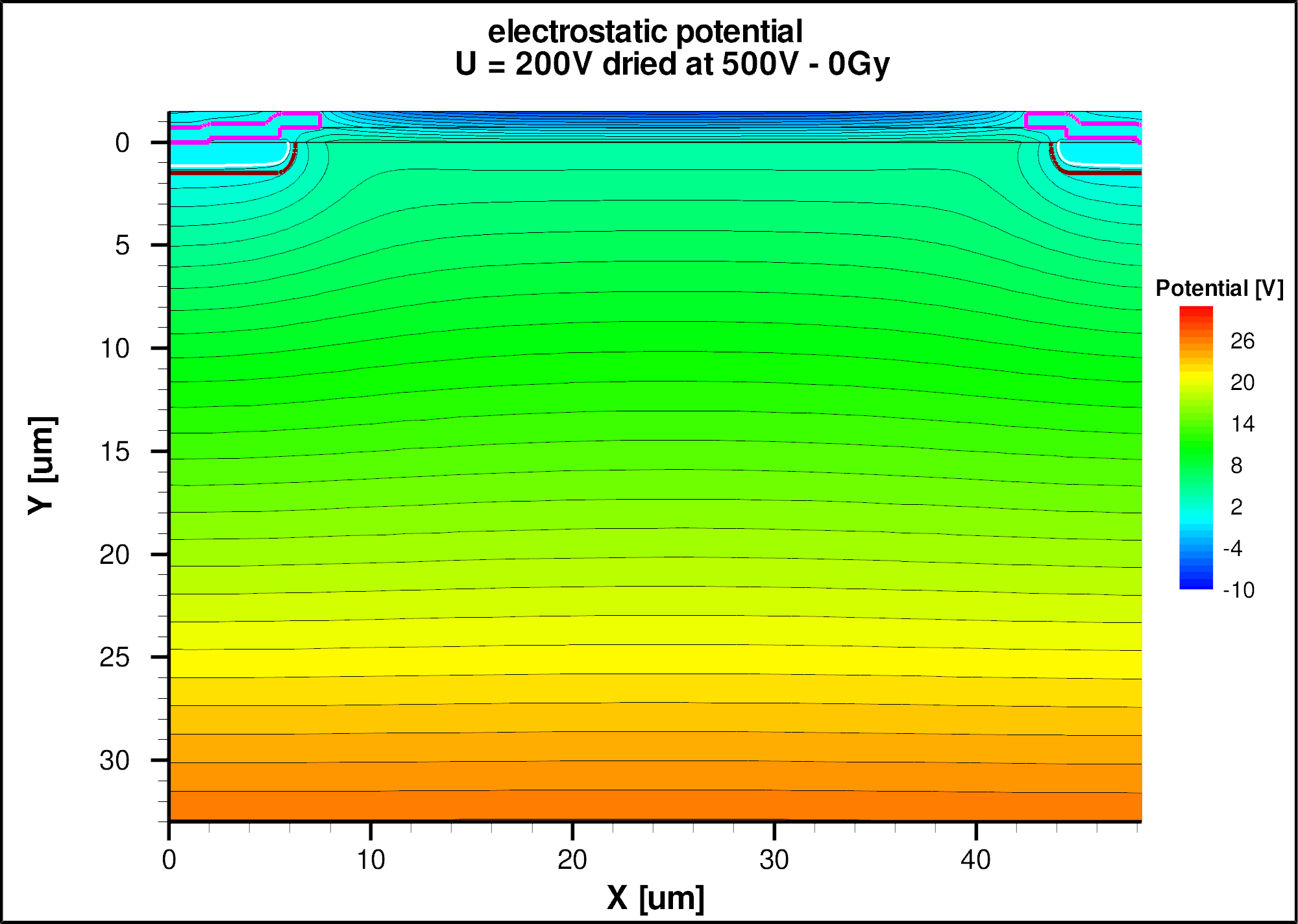}
   \caption{Simulated potential distributions for the non-irradiated sensor biased to 200~V for different biasing conditions. Left: "dried at 0~V. Right: "dried at 500~V". See Figure \ref{fig:sensor3} for the coordinate system.}
  \label{fig:potentials3}
 \end{figure}

    \subsection{Explanation of the charge losses} \label{sec:explain_losses}

 A detailed discussion and an explanation of the dependence of the charge losses on X-ray dose and biasing history has been presented in~\cite{Poehlsen:2012}. It is briefly summarised here.
 Figure~\ref{fig:potential}, taken from~\cite{Poehlsen:2012}, shows the simulated potential distribution for "humid", the situation where no charge losses are observed.
  In the calculations a positive oxide-charge density of $10^{11}$~cm$^{-3}$ for a non-irradiated sensor has been assumed.
 At the Si-SiO$_2$ interface the potential has a parabolic shape in the $x$~direction, with a maximum value of $\sim $~10~V in the center between the $p^+$~strips.
   In the $y$~direction the potential increases.
  Thus, for $eh$~pairs produced by the laser close to the interface, the electrons drift in the $y$~direction to the rear contact, the holes along the $x$~direction to the $p^+$~strips, and no charges are lost.

 The simulated potential for the condition "dried~@~0~V", where electrons are lost, is shown on the left of Figure~\ref{fig:potentials3}, again taken from~\cite{Poehlsen:2012}.
  For this condition the charge on top of the SiO$_2$~layer remains approximately zero, as it has been for the sensor at zero volt in steady-state conditions.
   The positive oxide charges cause an electron-accumulation layer at the Si-SiO$_2$~interface at a value of the potential of $\sim $~29~V, a saddle point of the potential $\sim $~5~$\upmu$m below the interface, and an electric field pointing from the SiO$_2$ into the silicon.
 Thus electrons drift towards the Si-SiO$_2$~interface where they are lost, i.e. not collected in the time interval during which the induced current is integrated.

  The right side of Figure~\ref{fig:potentials3} shows the potential for the condition "dried~@~500~V".
 In the humid steady-state condition at 500~V negative charges accumulate on top of the SiO$_2$.
  When the voltage is reduced to 200~V in dry conditions, the negative surface charges remain, overcompensate the positive oxide charges, produce a hole-accumulation layer at a potential value of $\sim $~4~V  at the Si-SiO$_2$~interface and an electric field which points from the silicon into the SiO$_2$.
 Thus holes drift to the Si-SiO$_2$~interface and are lost.

    \subsection{Explanation of the change of the charge losses} \label{sec:explain_change}

 Next we give a qualitative explanation of the change of the charge losses as function of pulse number and recovery time for the non-irradiated sensor.

  As seen in Figures~\ref{fig:0Gy_pulse} and \ref{fig:0Gy_time}, if no charges are lost, i.e. all charges are collected before the next laser pulse arrives, the charge losses remain zero.
   This is expected, as the conditions do not change from pulse to pulse.

 If, as for the situation "dried~@~500~V", positive charges are trapped close to the interface, the value of the potential at the interface in-between the $p^+$ strips will increase and approach the no-charge-loss situation shown in Figure~\ref{fig:potential}.
  As summarised in Table~\ref{tab:pulse_number}, after $\sim $~8~pulses of  $\sim $~130~000~$eh$~pairs spaced by 50~ns, the initial hole losses of $\sim 70$~\% have decreased to a saturation value of $\sim 25$~\%.
 We conclude that after $\sim $~8~pulses, the additionally trapped charges move away from the position where they were produced in the 50~ns time interval between the laser pulses.
  For the recovery of the charge losses Figure~\ref{fig:0Gy_time} shows a fast increase in the first few microseconds, followed by a much slower increase. The full recovery is reached at $\Delta t \approx 500$~$\upmu$s.
 We assume that the recovery is due to the diffusion of the excess holes over the potential barrier.

  Qualitatively the observations for the electron and for the hole losses are similar.
 The main difference is that the initial losses are $\sim 35$~\% for electrons, compared to $\sim 70$~\% for holes.
 Comparing the potential distributions shown in Figure~\ref{fig:potentials3}, bigger charge losses are expected for holes than for electrons. The electric field responsible for hole trapping (right) is higher and extends over a larger region than the one for electron trapping (left).
  As in the case of hole trapping, trapped electrons change the potential towards the zero-loss situation.
 However, trapped electrons reduce the value of the potential at the interface, whereas trapped holes increase it.

 Next we discuss the results for the irradiated sensor.
  The X-ray irradiation to 1~MGy increases the oxide-charge and interface-trap densities to an effective positive oxide-charge density of $\sim $~$2 \cdot 10^{12}$~cm$^{-2}$~\cite{Zhang:2011a}.
 In addition, the surface current increases by several orders of magnitude due to the interface states.

  Figure  \ref{fig:1MGy_pulse} shows, that for the condition "dried~at~500~V" no charge losses are observed when $\sim $~130~000~$eh$~pairs are generated per pulse.
 Apparently the negative surface charges on top of the passivation layer compensate the high effective oxide-charge density.
  For "humid" the density of negative surface charges is smaller, and does not fully compensate the positive oxide charges, and electron losses of $\sim 40$~\% for the first pulse are observed.
 For "dried~@~0~V" the surface-charge density is essentially zero, resulting in even higher electron losses of $\sim 70$~\% for the first pulse.

  The electron losses as a function of pulse number for the irradiated sensor behave quite differently than for the non-irradiated sensor.
 For "humid" and 130~000~$eh$~pairs generated per pulse, the electron losses decrease to essentially zero after $\sim 15$~pulses, whereas for "dried~@~0~V" they hardly decrease and show no sign of saturation.
  As seen in Figure~\ref{fig:1MGy_intensity}, a much higher number of $eh$~pairs is required for the irradiated sensor in the condition "dried~@~0~V" to significantly change the electron losses.
 Also the shape of the recovery of the electron losses, shown in Figure~\ref{fig:1MGy_time} for the irradiated sensor in conditions "humid", is different.
  Whereas for the non-irradiated sensor an initial partial recovery with time constants of less than 1~$\upmu$s is followed by a slow full recovery until $\sim $~200~$\upmu$s, 
	the electron losses for the irradiated sensor recover with a single time constant of $\sim $~6~$\upmu$s. We note, that in the discription by Equation (\ref{N_lost_fit}), $p \approx 1$ we interprete as a single time constant, and $p<1$ we interprete as a recovery with both, slower and faster components (compare Figures \ref{fig:0Gy_time} and \ref{fig:1MGy_time}).

  We finally comment, that we have made no attempt to simulate the dependence of the charge losses for the pulse structure used in the experiments.
  Given that it is a 3-D problem with charges spreading over large distances in-between the $p^+$ strips, a realistic simulation appeared out of reach.


     \subsection{Discussion of charge losses for high intensities } \label{sec:discuss_intensity}

 To further study the dependence of the electron losses on pulse number for the irradiated sensor in the condition "dried~@~0~V", the number of $eh$~pairs generated per pulse was varied between $10^5$ and $10^7$.
  The results are shown in Figure~\ref{fig:1MGy_intensity}.
 It is observed that  the fraction of electrons lost for the first pulse decreases from $\sim 70$~\% for $10^5$ to $\sim 20$~\% for $10^7$ $eh$~pairs.
  The explanation for this dependence is, that the charges deposited in a given pulse already change the local electric field, and thus already influence the charge collection for this first pulse.
 It is also observed that for $\gtrsim 3.6 \cdot 10^5$~$eh$~pairs generated, the electron losses saturate for higher pulse numbers. The saturation value decreases from $\sim 30$~\% for $3.6 \cdot 10^5$ to $\sim 2$~\% for $10^7$.
  We interpret this as evidence, that for the high radiation-induced effective oxide charge density and essentially zero negative charge on top of the SiO$_2$~layer, the maximum value of the potential at the Si-SiO$_2$~interface is high and many electrons have to be trapped to significantly reduce the electron losses.
 From the decrease of the charge losses for the first pulse with $eh$~intensity, we estimate that of the order of $10^6$~electrons have to be trapped locally in order to reduce the electron losses by about a factor 2.
  This number is significantly higher than for the irradiated sensor in conditions "humid", where already electron losses of $\sim 10^5$ make a significant difference, or for the electron and hole losses for the non-irradiated sensor.

    \section{Summary}

 Using the multi-channel Transient Current Technique, the currents induced by electron-hole pairs, produced by a focussed sub-nanosecond laser of 660~nm wavelength close to the Si-SiO$_2$ interface of $p^+n$-silicon strip sensors, have been measured, and charge-collection efficiencies determined.
  Sensors, before and after irradiation by 1~MGy (SiO$_2$) X-rays, have been investigated.

 For high densities of  electron-hole pairs deposited close to the Si-SiO$_2$ interface the plasma effect results in a significant increase in pulse length.
  However, the number of X-rays required to generate charge densities in this region so that these effects become significant are too high, to be of relevance for the AGIPD detector at the European XFEL.

 As already reported previously, dependent on radiation dose and biasing history, not all electrons or holes are collected at the contacts of the sensors within the typical readout integration times of order $\lesssim 100$~ns, but are trapped close to the Si-SiO$_2$ interface.
  These lost charges result in a non-steady state of the accumulation layers and the nearby electric fields, which causes a reduction of the charge losses.
 The number of trapped charges required to significantly reduce further charge losses and possibly reach constant values, varies between $\sim$~$10^5$ and $\sim$~$10^6$ in the investigated cases.
   The recovery times to steady-state conditions depends on the X-ray dose with which the sensor had been irradiated.

 Qualitative explanations of the findings have been given.
  Even if the results presented may be of limited practical relevance for the user of silicon sensors,  they provide further insight into the complexities of the Si-SiO$_2$-interface region of segmented $p^+n$-silicon sensors.

\section*{Acknowledgements}

  This work was performed within the AGIPD Project which is partially supported by the European XFEL-Company.
   We would like to thank the AGIPD colleagues for the excellent collaboration.
  Support was also provided by the Helmholtz Alliance "Physics at the Terascale", and the German Ministry of Science, BMBF, through the Forschungsschwerpunkt "Particle Physics with the CMS-Experiment".
   J.~Zhang was supported by the Marie Curie Initial Training Network "MC-PAD".


\section*{References}

\begin{thebibliography}{9}


 \bibitem{XFEL}
  M.~Altarelli et al. (Eds.),
   \emph{XFEL: The European X-Ray Free-Electron Laser, Technical Design Report}, Preprint DESY 2006-097, DESY, Hamburg 2006, ISBN 978-3-935702-17-1.

 \bibitem{XFEL2}
\url{http://www.xfel.eu/}.

 \bibitem{Tschentscher:2011}
   Th.~Tschentscher et al., TECHNICAL NOTE XFEL.EU TN-2011-001 ~2011,
   DOI:~10.3204/XFEL.EU/TR-2011-001.

 \bibitem{Graafsma:2009}
   H. Graafsma, 2009~JINST~4~P12011~2011,
    DOI:~10.1088/1748-0221/4/12/P12011.

 \bibitem{Klanner:2011}
  R.~Klanner et al.,
   \emph{Challenges for Silicon Pixel Sensors at the European XFEL},
    submitted to Nucl. Instr. and Meth. A, and arXiv~1212.5045.

 \bibitem{Becker:2010}
  J. Becker et al., Nucl. Instr. and Meth. A~615~(2010)~230-236,
   DOI:~10.1016/j.nima.2010.01.082.

 \bibitem{Becker:Thesis}
  J.~Becker, \emph{Signal development in silicon sensors used for radiation detection}, PhD thesis, Universit\"at Hamburg, DESY-THESIS-2010-33~(2010).

 \bibitem{Tove:1967}
   P.A.~Tove and W.~Seibt,  Nucl. Instr. and Meth. 51~(1967)~261.

 \bibitem{Poehlsen:2012}
  T. Poehlsen, et al.,
   Nucl. Instr. and Meth.~A~700~(2013) 22-39,
    DOI:~10.1016/j.nima.2012.10.063.

 \bibitem{Poehlsen1:2012}
  T. Poehlsen, et al.,
  \emph{Time dependence of charge losses at the Si-SiO$_2$ interface in $p^+n$-silicon strip sensors},
   submitted to Nucl. Instr. and Meth.~A.

 \bibitem{AGIPD}
  B.~Henrich et al.,
   Nucl. Instr. and Meth. A~500~Suppl.~1(2011)~S11, DOI:~10.1016/j.nima.2010.06.107.
	
\bibitem{AGIPD2}
	\url{http://photon-science.desy.de/research/technical_groups/detectors/projects/agipd/index_eng.html}.

 \bibitem{Hamamatsu}
  \url{http://www.hamamatsu.com/}.

 \bibitem{Zhang:2011a}
  J.~Zhang et al.,
  Journal of Synchrotron Radiation, 19~(2012)~340,
  DOI:~10.1107/S0909049512002348.

 \bibitem{Zhang:2011b}
 J.~Zhang et al.,
    JINST~6~C11013~(2011),
    DOI:~10.1088/1748-0221/6/11/C11013.

 \bibitem{Perrey:Thesis}
  H.~Perrey,
   \emph{Jets at Low Q$^2$ at HERA and Radiation Damage Studies for Silicon Sensors for the XFEL},
    PhD thesis, Universit\"at Hamburg, DESY-THESIS-2011-021 (2011).


 \bibitem{Kraner:1993}
  H.W.~Kraner, Z.~Li and E.~Fretwurst,
  Nucl. Instr. and Meth. A~326~(1993)~350.


















\end{thebibliography}
\end{document}